\documentclass[traditabstract]{aa}
\usepackage{graphicx}
\usepackage{txfonts}
\usepackage{wasysym}
\usepackage{natbib}

\begin{document}

\title{Accretion and structure of radiating disks}

\author{Patryk Mach and Edward Malec}
\institute{M. Smoluchowski Institute of Physics, Jagiellonian University, Reymonta 4, 30-059 Krak\'ow, Poland}

\date{}


\abstract{
We studied  a steadily accreting, geometrically thick disk model that selfconsistently takes into account selfgravitation of the polytropic gas, its interaction with the radiation and the mass accretion rate. The accreting mass is injected inward in the vicinity of  the central  $z=0$ plane, where also   radiation is assumed to be created. The rest  of the disk remains approximately stationary.  Only conservation laws are employed and the gas--radiation  interaction in the bulk of the disk is described in the thin-gas approximation.  We demonstrate that this scheme  is numerically viable and yields a  structure of the bulk that is influenced by the radiation and (indirectly) by the prescribed mass accretion rate. The obtained disk configurations are typical for environments in Active Galactic Nuclei (AGN), with the central mass of the order of $10^7 M_{\astrosun}$ to $10^8 M_{\astrosun}$, quasi-Keplerian rotation curves, disk masses ranging from about $10^6 M_{\astrosun}$ to $10^7 M_{\astrosun}$, and the luminosity ranging from $10^6 L_{\astrosun}$ to $10^9 L_{\astrosun}$. These luminosities are much lower than the corresponding Eddington limit.}

\maketitle

\section{Introduction}

There is a consensus   that  a standard model does exist  for geometrically thin  accretion disks  \citep[see e.g.][]{ShSu, LBP,pringle_81}.   Their structure is obtained assuming steady accretion and hydrodynamic equilibrium. While  the viscosity is needed to ensure accretion, it  can be completely eliminated in favour of the mass accretion rate -- once the steady disk solution is known to exist -- from the explicit formulation of structure equations. The  luminosity can be inferred from the disk geometry and the  mass accretion rate, but it does not influence the structure of steadily accreting disks. 

In contrast  to that, there is no generally accepted model for geometrically thick disks. We review just a sample of the related literature.    \citet{Paczynski78}  did not solve the full system of thick-disk  equations, but  assumed  elements of thin-disk approximation in determining the vertical structure.  \citet{Wiita} imposed some ingredients of thin-disk models onto thick disks. The luminosity was obtained analogously to thin disks -- it was deduced from the already known  disk structure.  In a similar analysis \citet{abramowicz_calvani_nobili_1980} assumed test gas approximation and  adapted the  mass accretion rate in a   way suitable  to ensure the energy conservation. In a  scenario  discussed in \citet{paczynski_abramowicz_1982} the mass accretion rate has been given as part of primary data and the bulk of the disk was convective. This description has been applied to supercritical accretion, but the selfgravity of the disk was not included. \citet{hashimoto_et_al_1993} and \citet{hashimoto_et_al_1995} considered optically and geometrically thick toroidal stars and Keplerian disks, with the selfgravity taken into account. They observed that the luminosity can be significantly higher than the Eddington limit. In these constructions the mass accretion rate has been estimated a posteriori.

The main goal of this investigation is to find a simple selfconsistent model of steady accretion with radiation and to study its structure.  We considered a scenario with a geometrically thick selfgravitating Newtonian  disk \citep[as in][]{hashimoto_et_al_1995}, in which the mass accretion rate is prescribed a priori \citep[as in][]{paczynski_abramowicz_1982}.  We assumed, following \citet{paczynski_abramowicz_1982}, that matter spirals  inwards in the immediate vicinity of the central disk plane $z=0$, where radiation is generated, by a phenomenological mechanism that  is inessential for the purpose of this analysis. This radiation interacts with the gas in the bulk of the disk   through Thompson scattering (i.e., we employed the thin-gas approximation).  The structure of the bulk is then obtained from the hydrodynamic equations (that include a radiation force) supplemented by the energy conservation equation.  We demonstrate that a self-contained description of accretion disks emerges -- a simple variant of radiation hydrodynamics. The whole problem is then analysed numerically, without any other simplifications.

 Our approach is inspired by the classical accretion model of \citet{Bondi}, in which one prescribes the mass accretion rate $\dot M$.  The mass accretion rate is not arbitrary, it has to agree with other accretion data: the asymptotic mass density $\rho_\infty$, the speed of sound $a_\infty$ and the equation of state. In the original model of Bondi the selfgravity of gas has been ignored,  but an analogous construction gives a model that also includes selfgravity  \citep{PRD2006}. As a consequence, the structure of the accreting spherical ball of fluid depends on $\dot M$. It has recently been discovered that these models are stable, even if selfgravity is taken into account \citep{MM08}.  
 
The content of the remainder of this paper is following. The notation is explained and the model is formulated in Sec.~2. Sec.~3 deals with radiating test fluids. We explicitly show that the equations can be solved by a Born-type approximation scheme, which appears to be convergent when the mass transfer rate is sufficiently low. Sec.~4.1 describes an iterative numerical scheme that has been specifically invented to solve this problem. Sec.~4.2 presents the numerical results.  The imposed conditions -- thin-gas approximation, steady accretion and approximately stationary disks -- appear quite stringent. We find disk configurations with the central mass of the order of $10^7 M_{\astrosun}$,  Obtained disks are geometrically thick, and become even thicker for higher accretion rates, but the effect of increasing accretion rate and radiation is not very strong. This is different from features known in ``Polish donuts'' \citep{abramowicz_jaroszynski_sikora_1978, abramowicz_calvani_nobili_1980, Wiita, paczynski_abramowicz_1982}. There is an upper limit for the luminosity, and it is significantly lower than the Eddington luminosity. These numerical solutions pass the recently formulated virial test described in \citet{mach_virial}.

\section{Description of the model}

\subsection{Notation and equations}

We assume that a steady accretion disk rotates around a central mass $M_\mathrm{c}$. The fluid is barotropic, i.e., $p=p(\rho)$, where $p$ is the pressure and $\rho$ is the mass density. Symbols $\mathbf U$ and $\Phi$ will be used to denote the velocity of the fluid and the gravitational potential, respectively.

We assume that there exists a phenomenological mechanism that produces radiation in the zone occupied by a radial infall flow (this coincides with the support of the mass accretion function $\dot M$ -- see a definition in Sec.~2.2), near the $z = 0$ plane, but we do not specify this mechanism. Pringle remarked that
 \textit{a steady disk can be constructed for almost any combination of viscosity and radiation process} \citep{pringle_81}. We believe that one can find -- by trial and error -- a suitable (point-dependent)  viscosity near the surface $z=0$ that can produce the radiation that is obtained in the way described below.

The radiation is produced with the emissivity per unit mass $j_\nu$ and undergoes the Thompson scattering. Our description of the radiative transfer follows \citet{padmanabhan}. Neglecting the change of the frequency of the scattered photon, we can write the scattering cross section as
\[ \frac{d \sigma \left(\hat \mathbf k \to \hat \mathbf k^\prime \right)}{d \Omega} = \sigma \varphi \left( \hat \mathbf k, \hat \mathbf k^\prime \right), \]
where $\sigma$ denotes the total Thompson scattering cross section, $\varphi$ describes the angular dependence, and
\[ \int d \Omega \varphi \left( \hat \mathbf k, \hat \mathbf k^\prime \right) = 1. \]
Here $\hat \mathbf k$ and $\hat \mathbf k^\prime$ denote appropriate directions in which the radiation propagates. 

The radiation transport can be described in terms of its intensity $I_\nu$. For a stationary process we have
\begin{equation}
\label{radiative_eq}
\hat \mathbf k \cdot \nabla I_\nu = \frac{\rho j_\nu}{4 \pi} - c \rho  \kappa I_\nu + c \rho \kappa \int d \Omega^\prime \varphi \left( \hat \mathbf k, \hat \mathbf k^\prime \right) I_\nu \left( \hat \mathbf k^\prime \right).
\end{equation}
Here $\kappa$ is the scattering opacity divided by the speed of light $c$. For the fully ionised hydrogen $\kappa = \sigma / (m_\mathrm{p} c)$.

Introducing the radiative flux
\[ \mathbf F_\nu = \int d \Omega \hat \mathbf k I_\nu \left( \hat \mathbf k \right) \]
and integrating Eq.~(\ref{radiative_eq}) over the solid angle, we obtain $\nabla \mathbf F_\nu = \rho j_\nu$, where we assumed that $j_\nu$ is independent of the direction $\hat \mathbf k$. The scattering terms that are present in Eq.~(\ref{radiative_eq}) cancel after integration.

In the following we will use the frequency integrated radiative flux or the radiation momentum density
\[ \mathbf j = \int d \nu \mathbf F_\nu, \]
so that
\begin{equation}
\label{div_j_eq}
\nabla \mathbf j = \rho \int d \nu j_\nu.
\end{equation}

The density and the velocity of the fluid must obey the continuity equation
\begin{equation}
\nabla \left( \rho \mathbf U \right) = 0.
\label{ab}
\end{equation}

The momentum conservation -- stationary Euler equations with the radiation term -- is given by
\begin{equation}
(\mathbf U \cdot \nabla) \mathbf U = -\nabla \Phi - \frac{1}{\rho} \nabla p + \kappa \mathbf j,
\label{aa}
\end{equation}
where $\Phi$ is the gravitational potential. The last term describes the interaction of gas and radiation in the thin-gas approximation; again, only Thomson scattering is taken into account.

The energy conservation equation states that the radiation energy flux $\mathbf j$ is correlated with the energy flux of the infalling gas:
\begin{equation}
\nabla \cdot \left( \mathbf U \rho \left( h + \frac{1}{2} \mathbf U^2 + \Phi \right) \right) + \nabla \cdot \mathbf j = 0,
\label{ad}					 
\end{equation}
where $h$ is the specific gas enthalpy ($dh = dp /\rho$).

Finally, the gravitational potential is given by the Poisson equation
\begin{equation}
\Delta \Phi = 4 \pi  G\rho,
\label{ac}
\end{equation}
where $G$ is the gravitational constant.

It is now clear that Eqs.~(\ref{ab})--(\ref{ac}) decouple from Eq.~(\ref{div_j_eq}) and any other possible equations describing the radiation transport.
 
Let $L$ denote the total luminosity: $L = \int_S d \mathbf S \cdot \mathbf j$, where $S$ is the surface of a disk and $d \mathbf S$ denotes an oriented normal surface element. Integration of Eq.~(\ref{ad}) over the disk volume leads to the approximate expression (notice that the neglected term with the enthalpy is comparatively small on the boundary)
\begin{equation}
L = - \int_S  d \mathbf S \cdot \mathbf U \rho \left( \frac{1}{2} \mathbf U^2 + \Phi \right).
\label{af}					 
\end{equation}
The luminosity depends on the shape of the disk, boundary values of the gravitational potential, rotation velocity, and the mass accretion rate. All these quantities are intricately related according to the differential equations; the shape itself can be determined a posteriori, after solving all of them.

\subsection{Axially symmetric equations}

We will seek axially symmetric solutions of Eqs.~(\ref{ab})--(\ref{ac}). The adopted coordinates are cylindrical variables $(r, \phi, z)$. Here $r$ is the cylindrical radius, while $R$ denotes the polar radius, so that $R = \sqrt{r^2 + z^2}$.

The condition of stationarity essentially implies $U^z = 0$; we assume that from now on. The velocity vector can be written as $\mathbf U = U \partial_r + \omega \partial_\phi$, where we shall demand $|U| \ll \omega r$.

Let us define the mass accretion function
\begin{equation}
\dot M = - 2 \pi U r \rho.
\label{bt}
\end{equation}
Eq.~(\ref{ab}) implies that $\dot M$ depends only on the variable $z$, i.e., $\dot M = \dot M(z)$.

The Euler equations read
\begin{eqnarray}
\partial_r \Phi + \frac{1}{\rho} \partial_r p + U \partial_r U - \omega^2 r & = & \kappa j^r, \nonumber \\
\partial_z \Phi + \frac{1}{\rho} \partial_z p  & = & \kappa j^z, \nonumber \\
U \left( \partial_r (r \omega) + \omega \right) & = & \kappa j^\phi. 
\label{ag} 
\end{eqnarray}     
For barotropes we have $\nabla p / \rho = \nabla h$, so that $\nabla (h + \Phi) = - (\mathbf U \cdot \nabla) \mathbf U + \kappa \mathbf j$. The expression on the right-hand side has a vanishing curl. Writing this explicitly, one finds that the consistency condition is $\partial_z \left( \omega^2 r \right) -\partial_z \left( U\partial_rU\right) = -\kappa \left( \partial_z j^r - \partial_r j^z \right)$. We search for solutions with small $U$, and therefore we neglect the term $\partial_z \left( U\partial_rU\right) $. If in addition the rotation velocity $\omega $ depends only on the radius, i.e., $\omega = \omega(r)$, then $\partial_z j^r - \partial_r j^z = 0$, and we can assume that $j^r = \partial_r \Psi$, $j^z = \partial_z \Psi$, where $\Psi(r,z)$ is a scalar function.   It is also convenient to introduce $\hat \Psi = \kappa \Psi$. We shall refer to $\Psi $ (or $\hat \Psi $) as to the radiation potential.
 
Let us take the full divergence of both sides of Eqs.~(\ref{ag}). After trivial rearrangements and neglecting the small term with $U$ one obtains
\begin{equation}
\Delta \left( h +  \Phi - \hat \Psi \right) = \frac{1}{r} \partial_r \left( \omega^2 r^2 \right). 
\label{ah}
\end{equation}
Introducing the centrifugal potential
\[ \Phi_\mathrm{c} = - \int^r dr^\prime r^\prime \omega^2(r^\prime), \]
and integrating equation (\ref{ah}), we obtain
\begin{equation}
h + \Phi + \Phi_\mathrm{c} - \hat \Psi = C,
\label{aj}
\end{equation}
where $C$ is a constant. Eq.~(\ref{aj}) can be also obtained directly from the Euler equations (\ref{aa}). From Eq.~(\ref{ad}) we have, again neglecting terms with $U$,
\begin{equation}
\Delta \hat \Psi  = \frac{\kappa \dot M }{2 \pi r} \partial_r \left( h + \Phi + \frac{1}{r} \omega^2 r^2 \right).
\label{ak}
\end{equation}
Combining Eqs.~(\ref{aj}) and (\ref{ak}), one arrives at
\begin{eqnarray}
\Delta \hat \Psi & = & \frac{\kappa \dot M}{2 \pi r} \partial_r \left( - \Phi_\mathrm{c} + \frac{1}{2} \omega^2 r^2 + \hat \Psi \right) \nonumber \\
& = & \frac{\kappa \dot M}{2 \pi r} \left( \partial_r \hat \Psi + 2 r \omega^2 + r^2 \omega \partial_r \omega \right).
\label{al}
\end{eqnarray}
We have to keep in mind that the right-hand side should be evaluated only in the region occupied by the disk. This is a linear equation that can be solved iteratively. There exists a unique solution that is regular in the open space $\mathbb R^3$.

We observe that the whole problem reduces to Eqs.~(\ref{ac}), (\ref{aj}) and (\ref{al}). Eqs.~(\ref{ac}) and (\ref{aj}) can be transformed into a nonlinear integral equation. The gravitational potential $\Phi$ can be expressed using the Green function formula
\begin{equation}
\Phi (\mathbf x) = - \frac{G M_\mathrm{c}}{R} + 4 \pi G \int_V d^3x^\prime \tilde G (\mathbf x - \mathbf x^\prime) \rho (\mathbf x^\prime),  
\label{am}
\end{equation}
where $-G M_\mathrm{c} / R$ is the contribution due to the central mass, $V \subset \mathbb R^3$ is the region occupied by the disk and $\tilde G$ denotes the Green function of the laplacian in the open space $\mathbb R^3$, i.e., $\tilde G(\mathbf x) = - 1/(4 \pi | \mathbf x |)$. Eq.~(\ref{am}) can be combined with Eq.~(\ref{aj}), and it is convenient to exploit the fact that the density $\rho$ is connected with the specific enthalpy $h$ by the assumed equation of state. For the polytropic equation of state $p = K \rho^\Gamma$ the specific enthalpy is $h = (K \Gamma / (\Gamma - 1)) \rho^{\Gamma - 1}$, and we obtain
\begin{eqnarray}
h & - & \frac{G M_\mathrm{c}}{R} + \Phi_\mathrm{c} - \hat \Psi \nonumber  \\
& + & 4 \pi G \left( \frac{\Gamma - 1}{K \Gamma} \right)^\frac{1}{\Gamma-1} \int_V d^3 x^\prime \tilde G (\mathbf x - \mathbf x^\prime) h^\frac{1}{\Gamma -1}(\mathbf x^\prime) = C.
\label{ao}
\end{eqnarray}

In summary, we need the following elements to describe the hydrostationary equilibrium of a Newtonian radiating disk: (i) the accretion mass function $\dot M(z)$, the rotation curve $\omega(r)$, the central gravitational potential $-G M_\mathrm{c} / R$ and the equation of state $p = p(\rho)$; (ii) the radiation potential $\hat \Psi$, which can be determined from the linear Eq.~(\ref{al}). It is convenient to write it down using the integral equation
\[ \hat \Psi(\mathbf x) = \frac{1}{2 \pi} \int_V d^3 x^\prime \tilde G (\mathbf x - \mathbf x^\prime) \frac{\kappa \dot M}{ r^\prime} \partial_{r^\prime} \left( \frac{1}{2} \omega^2 {r^\prime}^2 - \Phi_\mathrm{c} + \hat \Psi \right). \]
(iii) The distribution of the density or the specific enthalpy. This can be obtained from Eqs.~(\ref{ac}) and (\ref{aj}), or from an equation similar to Eq.~(\ref{ao}).

In all above formulae the rotation law $\omega = \omega(r)$ is treated as known a priori. A popular choice for test fluid solutions is to assume a (modified) Keplerian rotation 
\begin{equation}
\omega^2 = \frac{G M_\mathrm{c}}{\left( r^2 + z_0^2 \right)^{3/2}},
\label{ar}
\end{equation}
where $z_0$ is a constant. In this case the centrifugal potential is $\Phi_\mathrm{c} = G M_\mathrm{c} / \sqrt{r^2 + z_0^2}$. Other simple possibilities are the rigid rotation $\omega = \mathrm{const}$, the rotation law $\omega = v_0/r$, where $v_0 = \mathrm{const}$ is the linear angular velocity, and $\omega = j_0/r^2$, where $j_0 = \mathrm{const}$ is the specific angular momentum. We will refer to the last relation as to the $j$-const rotation law.

The $j$-const rotation is exceptional. In this case we have
\[ 2 r \omega^2 + r^2 \omega \partial_r \omega = 0, \]
and the radiation equation takes the form
\[ \Delta \hat \Psi = \frac{\kappa \dot M}{2 \pi r} \partial_r \hat \Psi.  \]
Assuming that $\hat \Psi$ is constant at the spatial infinity, we can conclude that $\hat \Psi \equiv \mathrm{const}$ everywhere. For the proof, notice that it suffices to deal with the homogeneous case where $\hat \Psi \to 0$ as $R \to \infty$. Multiply the last equation by $\hat \Psi$ and integrate over $\mathbb R^3$. Integrating by parts and employing the fact that $\partial_r \dot M = 0$ one arrives at
\[ -\int_{\mathbb R^3} d^3 x \left( \nabla \hat \Psi \right)^2=   \kappa  \int_{-\infty }^{\infty } dz\dot M  \left( \hat \Psi \right)^2. \]
The left-hand side is nonpositive, while the right-hand side is nonnegative. Therefore $\hat \Psi = \mathrm{const}$. That is a torus with a $j$-const rotation law is not emitting any radiation in $r$- and $z$-directions. Since the $j^\phi$ component of the radiation flux vector vanishes identically for the $j$-const rotation, we conclude that this rotation law is not compatible with the radiation. This is consistent with the picture emerging from the analysis of \citet{LBP}; the radiated luminosity balances the energy budget of the accreting matter and it is accompanied by the shedding of the angular momentum.  

The quantity $C$ in Eq.~(\ref{aj}) is a free parameter. The boundary of the disk is defined as a closed two-surface on which the specific enthalpy vanishes; thus it cannot be defined a priori; it is free. The exception is the test gas approximation without radiation, where the shape of a boundary is completely dictated by the central potential and the rotation curve $\omega(r)$. For radiating disks, the shape of a boundary also depends on the luminosity, which in turn is related to the mass accretion rate. To uniquely define a disk, one needs additional information. These questions will be discussed in forthcoming sections.  
 
\subsection{Luminosity and fluxes}

Two of the flux densities are given by $j^r = \partial_r \Psi$, $j^z = \partial_z \Psi$. The third component $j^\phi $ can be obtained from the $\phi$-th Euler equation.

The formula (\ref{af}) yields for a disk with the minimal and maximal radial extensions $r_\mathrm{in} $ and   $r_\mathrm{out}$, respectively,
\begin{equation}
L \equiv \int_S d\mathbf S \cdot \mathbf j \approx \int dz \dot M \left( \frac{1}{2} \left( \omega r \right)^2 + \Phi \right)_{r_\mathrm{in}}^{r_\mathrm{out}},
\label{ap}
\end{equation}
where we employed the condition $|U| \ll \omega r$. This agrees with the formula derived by \citet{LBP} for the central  star that is co-rotating with the disk, if $r_\mathrm{in} \ll r_\mathrm{out}$. The local formulae for the flux densities are different. In our case the flux density $\mathbf j$ is defined uniquely, whereas in the standard approach it is given up to the total divergence \citep{LBP}. The accretion flow originates outside of the disk and falls onto the central body. It is concentrated close to the plane $z=0$. The quantity $\dot M$ does not depend on $r$ and thus the mass density cannot vanish at the boundary; that means that the actual shape of the disk is not well defined near $z=0$. Nevertheless, in formula (\ref{ap}) we assume that the disk extends from a definite exterior cylinder ($r_\mathrm{out}$) to a definite inner cylinder ($r_\mathrm{in}$).

\section{Test fluid solutions}

Assume that the gas density is low and the gravitational potential is dominated by the central term $- G M_\mathrm{c} / R^3$. It is reasonable to expect (and in fact this expectation can be proved) that there are solutions of Eq.~(\ref{ah}) that are well approximated by solutions of the linear inhomogeneous equation
\begin{equation}
\Delta h = \frac{1}{r} \partial_r \left( \omega^2 r^2 \right).
\label{aq}
\end{equation}
Consider the rotation law given by Eq.~(\ref{ar}). One can check that
\begin{equation}
h = G M_\mathrm{c}  \left( \frac{1}{R} - \frac{1}{\sqrt{r^2 + z_0^2}}\right) 
\label{as}
\end{equation}
solves Eq.~(\ref{aq}), with the boundary given by two planes  $|z| = z_0$. For small $z_0$ one recovers the solution of \citet{Paczynski78}:
\[ h =  G M_\mathrm{c} \left( \frac{1}{2 r^3} \left( z^2_0 - z^2 \right) \right). \]
Notice that the boundary is not closed. We show below that radiation causes the closure of the external end of the disk, but it is the influence of selfgravity that can make 
a finite disk.

\subsection{Radiation}

Let the mass of the disk be negligible and the potential be dominated by the central term $\Phi = -G M_\mathrm{c} / R$. Assume a modified Keplerian rotation curve (\ref{ar}). The solution of Eqs.~(\ref{ah}) and (\ref{al}) can be found perturbatively, treating the mass accretion term as a perturbation.
 
The zeroth-order term $h_0$ is given by Eq.~(\ref{as}), that is
\begin{equation}
h_0 = G M_\mathrm{c} \left( \frac{1}{R} - \frac{1}{\sqrt{r^2 + z_0^2}}\right),
\label{ca}
\end{equation}
and the $h_1$ term is
\begin{eqnarray}
h_1 \left( \mathbf x \right) & = &  \int_V d^3 x_1 \tilde G \left( \mathbf x_1 - \mathbf x \right)
\frac{\kappa \dot M}{2 \pi r_1} \partial_{r_1} \left( \Phi + h_0 + \frac{1}{2} \omega^2 r^2_1 \right) \nonumber \\
& = & - \int_V d^3 x_1 \tilde G \left( \mathbf x_1 - \mathbf x \right) \frac{\kappa \dot M}{4 \pi r_1} G M_\mathrm{c} \times \nonumber \\
& & \times \partial_{r_1} \left( \frac{1}{\sqrt{r_1^2 + z_0^2}} + \frac{z_0^2}{\left( r_1^2 + z_0^2 \right)^{3/2}}\right).
\label{aw}
\end{eqnarray}
Higher order terms $h_k$ ($k = 2, 3, \dots$) are defined by
\begin{equation}
h_k \left( \mathbf x \right) = \int_V d^3 x_k \tilde G \left( \mathbf x_k - \mathbf x \right) \frac{\kappa \dot M}{2 \pi r_k} \partial_{r_k} h_{k-1} \left( \mathbf x_k \right).
\label{ax}
\end{equation}
Here, for $k = 1, 2, \dots$, $\mathbf x_k = (r_k, z_k, \phi_k)$; $r_k$, $z_k$ and $\phi_k$ are the cylindrical components of $\mathbf x_k$, and the integration volume element $d^3 x_k$ equals $d^3 x_k = dr_k dz_k d \phi_k r_k$. The unlabelled quantities $r$, $z$ and $\phi$ are reserved herein and in what follows for the components of the vector $\mathbf x = (r, z, \phi)$.

The specific enthalpy corresponding to the full solution of Eqs.~(\ref{ah}) and (\ref{al}) is $h = h_0 + h_1 + \delta h$, where $\delta h = h_2 + h_3 + \dots$ It is easy to check that all individual functions $h_1$, $h_2$, $h_3$, \dots are non-positive. Assume furthermore that the disk extends from the inner cylinder of a radius $r_\mathrm{in}$ to an outer cylinder of a radius $r_\mathrm{out}$. Then it is a simple exercise to show that
\begin{equation}
|\partial_{r_k} h_k| \le \frac{|h_k|}{r_\mathrm{in}},
\label{ay}
\end{equation} 
for any $k = 1, 2,\dots$

Stationary disks can exist when the above expansion scheme is convergent. The application of  standard criteria for convergence of series leads to a bound onto the mass accretion rate $\dot M$ and on the total luminosity. These statements can be proved quite generally. We would like to point out that the rotation law (\ref{ar}) is not compatible with the geometry of a closed disk. While it allows for a closure at a finite distance from the centre, it should be modified in the interior region to give a disk configuration. Let us point out, however, that the inclusion of   self-gravity should allow for a closed disk configuration. 
 
\subsection{Radiation and the boundary of the disk}

The boundary of a disk is a smallest surface where the specific enthalpy vanishes: $h = h_0 + h_1 + \delta h \equiv 0$. In the absence of accretion, i.e., when $\dot M = 0$, we have $h_k = 0$ for $k = 1, 2, \dots$ Thus at $z = z_0$ there exists a horizontal part of the boundary.
   
If the mass accretion term $\dot M$ is positive, then it is clear from the inspection of Eqs.~(\ref{aw}) and (\ref{ax}) that $h_1$ and $\delta h$ are non-negative. This means that the boundary is pushed inward to a location with a value $|z| < z_0$.
 
The directional derivative of the specific enthalpy $h$ vanishes along the boundary: $dh = 0$, $dh = \partial_r h dr + \partial_z h dz$. Notice that $\kappa j_r= \partial_r \left( h_1 + \delta h \right)$ and $\kappa j_z = \partial_z \left( h_1 + \delta h \right)$. This leads to the equation 
\begin{equation}
\frac{dz}{dr} = - \frac{\kappa j_r + \partial_r h_0}{\kappa j_z + \partial_z h_0} = - \frac{\kappa j_r + \partial_r h_0}{\kappa j_z - \partial_z \Phi}.
\label{az}
\end{equation}
Values of $j_r$ and $j_z$ are positive on the upper face of the disk ($z>0$) and negative on the lower face ($z<0$). For vanishing radiation, Eq.~(\ref{az}) reduces to the well-known formula $dz/dr = \left( \omega^2 r - \partial_r \Phi \right) / \partial_z \Phi$.

\subsection{A family of analytic solutions}

Assume $\dot M \equiv F(r) \delta(z)$, where $F(r) \equiv F = \mathrm{const} >0$ for $r_\mathrm{in} \le r \le r_\mathrm{out}$ and $F(r) = 0$ for $r < r_\mathrm{in}$ and $r > r_\mathrm{out}$. The delta-like distribution can be easily handled analytically, and a close inspection shows that some aspects of the disk solutions weakly depend on the specific form of the mass accretion function $\dot M$. We assume that $F \ll \int dz \omega r^2 \rho$, because the radial drift should be negligible in comparison to the rotational motion. For $r_\mathrm{in} < b < r <r_\mathrm{out}$ the rotation curve is assumed to be given by Eq.~(\ref{ar}), which yields the $h_0$ enthalpy term according to Eq.~(\ref{ca}). With this assumption one can obtain the cusp-like end of the external part of the disk. This rotation law does not allow to close the internal part of the disk. To achieve this, we have to modify the rotation curve in a transient region just above $r_\mathrm{in}$. This transient zone is not particularly important, since it does not seriously impact the structure of the disk nor its luminosity. For instance, one can take in the interval $r_\mathrm{in} < r < b$,
\begin{equation}
\omega^2 = \frac{G M_\mathrm{c}}{\left( r^2 + z_0^2 \right)^{3/2}} \left( 2 + \left( \alpha - 1 + \alpha \left( \frac{z_0}{r} \right)^2 \right) \left(\frac{r}{b} \right)^\alpha \right).
\label{ba}
\end{equation}
The constant $F$ describes the intensity of the radial  baryonic drift, and it can be found from the condition that $h(r_\mathrm{out})=0$. The parameter $\alpha $ can be specified by the demand that $h(r_\mathrm{in}) = 0$. Notice that the rotation curve (\ref{ba}) yields the specific enthalpy
\[ h_0 = G M_\mathrm{c} \left( \frac{1}{R} - \frac{2 - (r/b)^\alpha}{\sqrt{r^2 + z_0^2}} \right) \]
in the region $(r_\mathrm{in}, b)$. It is continuous everywhere, albeit it is not differentiable at $r=b$.

Let us define the elliptic function
\begin{eqnarray*}
E \left( r, r_k, z \right) & = & - \int_0^{2 \pi} d \phi_k \left. \tilde G \left( \mathbf x - \mathbf x_k \right) \right|_{z_k = 0} \\
& = & \frac{1}{4 \pi} \int_0^{2 \pi} \frac{d \phi_k}{\sqrt{r^2 + r_k^2 + z^2 - 2 r_k r \cos \phi_k}}
\end{eqnarray*}
(here $k = 1, 2,\dots$). The term $h_1$ is now given by
\begin{eqnarray}
\label{bd}
h_1(r, z) & = & - \frac{\kappa F G M_\mathrm{c}}{4 \pi} \int_{r_\mathrm{in}}^{r_\mathrm{out}} dr_1 r_1 E \left( r, r_1, z \right) \times \\ 
& & \times \left( \frac{1}{\left( r_1^2 + z_0^2 \right)^{3/2}} + \frac{3 z_0^2}{\left( r_1^2 + z_0^2 \right)^{5/2}}\right) \nonumber \\
& & + \frac{\kappa F G M_\mathrm{c}}{2 \pi} \int_{r_\mathrm{in}}^{b} dr_1  E \left( r, r_1, z \right) \times \nonumber \\
& & \times \partial_{r_1} \left( \frac{1 - \left( 1 + \alpha \right) \left( \frac{r_1}{b} \right)^\alpha}{2 \left( r_1^2 + z_0^2 \right)^{1/2}} + \frac{z^2_0 \left( 1 - \left( \frac{r_1}{b} \right)^\alpha \right)}{2\left( r_1^2 + z_0^2 \right)^{3/2}} \right). \nonumber
\end{eqnarray}
One can specify the parameters $b$ and $\alpha$ of the transition so that the second integral in Eq.~(\ref{bd}) can be neglected, at least for thin disks.

The next expansion terms $h_k$'s ($k = 2, 3,\dots$) are given by  
\[ h_k(\mathbf x) = - \frac{\kappa F}{2 \pi} \int_{r_\mathrm{in}}^{r_\mathrm{out}} dr_k \left. E \left( r, r_k, z \right) \partial_{r_k}  h_{k-1} \right|_{z_k = 0}. \]
We can use estimate (\ref{ay}) to obtain the inequality
\begin{equation}
\sup |h_k| \le \frac{\kappa F}{2 \pi r_\mathrm{in}} \sup \left( \int_{r_\mathrm{in}}^{r_\mathrm{out}} dr_k  E \left( r, r_k, z \right) \right) \sup |h_{k-1}|.
\label{bf}
\end{equation}
The elliptic function $E \left( r, r_k, z \right)$ is integrable on any finite disk. The inequality (\ref{bf}) means that the series expansion is convergent for sufficiently low values of the product
\[ \frac{\kappa F}{2 \pi r_\mathrm{in}} \int_{r_\mathrm{in}}^{r_\mathrm{out}} dr_k  E \left( r, r_k, z \right). \]
It suffices that the mass current constant $F$ is small enough, which means -- notice the conservation law (\ref{ap}) -- that stationary disks cannot have arbitrarily high luminosity. 

The mass accretion rate influences the disk geometry, as seen from the following discussion. At the boundary we have $h_0 + h_1 \approx 0$; thus (keeping only the leading term of $h_1$) we have 
\begin{eqnarray}
\frac{1}{R} - \frac{1}{\sqrt{r^2 + z_0^2}} & = & \frac{\kappa F}{4 \pi} \int_{r_\mathrm{in}}^{r_\mathrm{out}} dr_1 r_1 E \left( r, r_1, z \right) \times \nonumber \\ 
& & \times \left( \frac{1}{\left( r_1^2 + z_0^2 \right)^{3/2}} + \frac{3 z_0^2}{\left( r_1^2 + z_0^2 \right)^{5/2}}\right)
\label{bh}
\end{eqnarray}
for $r>b$ and 
\begin{eqnarray}
\frac{1}{R} - \frac{2 - \left( r / b \right)^\alpha}{\sqrt{r^2 + z_0^2}} & = & \frac{\kappa F}{4 \pi} \int_{r_\mathrm{in}}^{r_\mathrm{out}} dr_1 r_1 E \left( r, r_1, z \right) \times \nonumber \\
& & \times \left( \frac{1}{ \left( r_1^2 + z_0^2 \right)^{3/2}} + \frac{3 z_0^2}{ \left( r_1^2 + z_0^2 \right)^{5/2}}\right)
\label{bi}
\end{eqnarray}
for $r\le b$. Assuming that the outer part of the disk extends up to $r_\mathrm{out}$, one obtains the value of $F$
\begin{equation}
\frac{\kappa F}{4 \pi} \approx \frac{z_0^2}{2 r_\mathrm{out} I},
\label{bj}
\end{equation}
where  
\[ I = \int_{r_\mathrm{in}/r_\mathrm{out}}^{1} dx x E \left( 1, x, 0 \right) \left( \frac{1}{ \left( x^2 + z_0^2 / r_\mathrm{out}^2 \right)^{3/2}} + \frac{3 z_0^2 / r_\mathrm{out}^2}{ \left( x^2 + z_0^2 / r_\mathrm{out}^2 \right)^{5/2}}\right). \]
One can check that
\begin{equation}
\label{inequality_i}
I \ge \int_{r_\mathrm{in}}^{r_\mathrm{out}} dr_k E \left( r_\mathrm{out}, r_k, 0 \right) \ge \gamma(r_\mathrm{in}/r_\mathrm{out}) \int_{r_\mathrm{in}}^{r_\mathrm{out}} dr_k E \left( r, r_k, z \right)
\end{equation}
for small $z_0$. Here $\gamma(r_\mathrm{in}/r_\mathrm{out})$ is a coefficient, with values depending on the ratio $r_\mathrm{in}/r_\mathrm{out}$, ranging between 0.01 and 1 for $r_\mathrm{in}/r_\mathrm{out} = 10^{-9}, \dots, 1$. It is found numerically as the largest coefficient ensuring that the inequality
\[ \int_{r_\mathrm{in}/r_\mathrm{out}}^1 dx E(1,x,0) \ge \gamma(r_\mathrm{in}/r_\mathrm{out}) \int_{r_\mathrm{in}/r_\mathrm{out}}^1 dx E(r/r_\mathrm{out},x,0) \]
is satisfied. Inequality~(\ref{inequality_i}) combined with Eq.~(\ref{bj}) yields
\[ \frac{\kappa F}{2 \pi r_\mathrm{in}} \int_{r_\mathrm{in}}^{r_\mathrm{out}} dr_k E \left( r, r_k, z \right) \le \frac{z_0^2}{\gamma(r_\mathrm{in}/r_\mathrm{out}) r_\mathrm{in} r_\mathrm{out}}. \]
Obviously, for thin disks the right-hand side is much smaller than one; this implies the convergence of our approximation scheme. The consistency condition $F\ll \int dz \omega r^2 \rho$, discussed earlier, can be checked only a posteriori, after solving all equations.

Eq.~(\ref{bi}) allows, putting $r = r_\mathrm{in}$, to specify the parameter $\alpha$ that appears in the rotation curve in the transient zone $(r_\mathrm{in}, b)$. By a proper choice of $b$, $r_\mathrm{in}$, $r_\mathrm{out}$ and $z_0$ one can always achieve $\alpha \ll 1$, which gives the total luminosity close to the value corresponding to the Keplerian rotation curve. Formula (\ref{ap}) now yields
\[ L \approx \frac{F G M_c}{2} \left( \frac{1}{r_\mathrm{in}} - \frac{1}{r_\mathrm{out}} \right). \]
Other quantities, including the gas density, can be found from the Euler equations and the radiation transport equations.  
\section{Heavy selfgravitating disks}

\subsection{Numerical methods}

Solutions corresponding to heavy selfgravitating disks can be obtained numerically by solving Eqs.~(\ref{ac}), (\ref{aj}) and (\ref{al}). In the following we assume the polytropic equation of state $p = K \rho^\Gamma$, so that $h = (K \Gamma / (\Gamma - 1)) \rho^{\Gamma - 1}$. The rotation law is fixed up to a multiplicative constant. Numerical examples will be given for Keplerian-type rotation  $\omega \sim r^{-3/2}$.


Assume that the disk spreads up to $R = R_\mathrm{out}$ at the equatorial plane, and that the maximal density of the gas is $\rho_\mathrm{max}$. The quantity $u = G R^2_\mathrm{out} \rho_\mathrm{max}$ has the dimension of the potentials $\Phi$, $\Phi_\mathrm{c}$ and $\hat \Psi$. It can be used to define following dimensionless quantities $\tilde \Phi = \Phi / u$, $\tilde \Phi_\mathrm{c} = \Phi_\mathrm{c} / u$, $\tilde \Psi = \hat \Psi / u$, $\tilde K = K \rho_\mathrm{max}^{\Gamma - 1} / u$, $\tilde \omega = \omega R_\mathrm{out}/\sqrt{u}$. In similar fashion we introduce $\tilde \rho = \rho / \rho_\mathrm{max}$ and $\tilde M_\mathrm{c} = M_\mathrm{c} / (\rho_\mathrm{max} R_\mathrm{out}^3)$. Dimensionless spatial coordinates are defined as $\tilde {\mathbf x} = \mathbf x / R_\mathrm{out}$. The relevant equations of the model can be rewritten in the form
\begin{equation}
\label{rescaled_a}
\tilde K \Gamma /(\Gamma - 1) \tilde \rho^{\Gamma - 1} + \tilde \Phi + \tilde \Phi_\mathrm{c} - \tilde \Psi = \tilde C,
\end{equation}
\begin{equation}
\label{rescaled_b}
\tilde \Delta \tilde \Phi = 4 \pi \tilde \rho,
\end{equation}
\begin{equation}
\label{rescaled_c}
\tilde \Delta \tilde \Psi = \tilde S,
\end{equation}
where $\tilde C = C / u$, $\tilde \Delta$ is the laplacian with respect to the rescaled coordinates $\tilde {\mathbf x}$, and
\[ \tilde S  =  \frac{\kappa \dot M}{2 \pi \tilde r} \left( \partial_{\tilde r} \tilde \Psi + 2 \tilde r \tilde \omega^2 + \tilde r^2 \tilde \omega \partial_{\tilde r} \tilde \omega \right). \]

The system of Eqs.~(\ref{rescaled_a}) and (\ref{rescaled_b}) with $\tilde \Psi = 0$, i.e., without radiation, corresponds to a simple model of a rotating star or a toroid. There is a long tradition in the numerical analysis of these equations \citep[see e.g.][]{stoeckly_1965, ostriker_mark_1968, eriguchi_1978, eriguchi_muller_1985}. In particular, it is known that they can be solved iteratively starting with some initial guess of the density distribution. The very fruitful iterative scheme of \citet{ostriker_mark_1968} based on the Green function expression for the gravitational potential is known as the self-consistent field method. It has been used successfully by many authors and in many variants, including computation of general-relativistic solutions \citep[cf.][]{clement_1974, blinnikov_1975, komatsu_eriguchi_hachisu_1989, nishida_eriguchi_lanza_1992}. An analytic expansion scheme has been also proposed by \citet{Odrzywolek}. Since in our case Eq.~(\ref{rescaled_c}) for the radiation potential is also Poisson-like, the numerical method of this paper follows the self-consitent field pattern.

Knowing the density distribution, one can compute the gravitational potential $\tilde \Phi$ from a rescaled version of Eq.~(\ref{am}), i.e., 
\begin{equation}
\label{rescaled_d}
\tilde \Phi (\tilde {\mathbf x}) = -\frac{\tilde M_\mathrm{c}}{\tilde R} + 4 \pi \int_V d^3 x \tilde G(\tilde{\mathbf x} - \mathbf x)\tilde \rho (\mathbf x).
\end{equation}
Eq.~(\ref{rescaled_d}) is clearly not suited for a direct numerical implementation due to the singularity at $\tilde {\mathbf x} = \mathbf x$ present in the Green function $\tilde G (\tilde {\mathbf x} - \mathbf x) = -1/(4 \pi |\tilde {\mathbf x} - \mathbf x|)$. This  difficulty can be avoided by a regularisation procedure. The   standard trick   is to expand  the Green function in terms of spherical harmonics. Assuming axial symmetry, and spherical coordinates $(\tilde R, \tilde \mu = \cos \tilde \theta, \tilde \phi)$ we can write
\begin{eqnarray}
\label{phi_series}
\tilde \Phi(\tilde R,\tilde \mu) & = &  -\frac{\tilde M_c}{\tilde R} - 4 \pi \sum_{j=0}^\infty P_{2j}(\tilde \mu) \int_0^\infty dR {R}^2 F_{2j}(\tilde R,R) \times \nonumber \\
& & \times \int_0^1 d \mu P_{2j}(\mu) \tilde \rho(R,\mu),
\end{eqnarray}
where
\[ F_j(\tilde R,R) = \left\{ \begin{array}{ll}
\frac{1}{\tilde R} \left( \frac{R}{\tilde R} \right)^j, & \tilde R \ge R, \\
\frac{1}{R} \left( \frac{\tilde R}{R} \right)^j, & \tilde R < R,
\end{array} \right. \]
and $P_j$ denotes the $j$-th Legendre polynomial. From the numerical point of view the regularisation of the discretised integral in expression (\ref{rescaled_d}) consists of taking a finite number of terms in the series in Eq.~(\ref{phi_series}).

The equation for the radiation potential $\tilde \Psi$ can be solved in a similar fashion. Assuming that $\tilde S$ is known, we can obtain $\tilde \Psi$ as
\begin{eqnarray}
\label{radiation_series}
\tilde \Psi(\tilde R,\tilde \mu) & = & - \sum_{j=0}^\infty P_{2j}(\tilde \mu) \int_0^\infty dR {R}^2 F_{2j}(\tilde R,R) \times \nonumber \\
& & \times \int_0^1 d \mu P_{2j}(\mu) \tilde S(R,\mu).
\end{eqnarray}
Knowing $\tilde \Psi$ is necessary to compute $\tilde S$, which in turn is needed to find $\tilde S$. We demonstrate that in the case investigated here an iterative process can be applied that eventually yields an appropriate solution both for $\tilde S$ and $\tilde \Psi$. The expression for $\tilde S$ is valid in the interior of  the  disk, i.e., where $\tilde \rho \neq 0$. The integral in Eq.~(\ref{radiation_series}) has to be evaluated only over this region, hence $\tilde \rho$ is implicitly assumed to be known.

Conversely, once $\tilde \Phi$ and $\tilde \Psi$ are known, we can compute the density $\tilde \rho$ directly from Eq.~(\ref{rescaled_a}). This equation, as it is written, allows negative values of enthalpy. In this case one has to modify the physical boundary of the disk. We will only search for $\tilde \rho$ in domains where the enthalpy given by Eq.~(\ref{rescaled_a}) is positive; otherwise we will assume $\tilde \rho = 0$.

The strategy of solving Eqs.~(\ref{rescaled_a}), (\ref{rescaled_b}) and (\ref{rescaled_c}) is as follows. We fix the value of the central mass $\tilde M_\mathrm{c}$, and the ratio of the inner and outer radii of the boundary of the disk $R_\mathrm{in} / R_\mathrm{out}$. In the next step we assume temporarily that $\tilde \Psi = 0$, and construct an initial density distribution in the form of a toroid with assumed ratio $R_\mathrm{in} / R_\mathrm{out}$ and the maximum density $\tilde \rho = 1$. For this toroid, the gravitational potential can be obtained form Eq.~(\ref{phi_series}), and the new density distribution can be computed from Eq.~(\ref{rescaled_a}). This procedure is iterated until a desired convergence is reached. After establishing a hydrostatic structure of the selfgravitating disk, we ``switch on'' the radiation. We fix the function $\kappa \dot M(\tilde z)$ and obtain the radiation potential coming from the already found hydrostatic structure. This can be performed by solving iteratively Eq.~(\ref{radiation_series}). Finally, all three Eqs.~(\ref{phi_series}), (\ref{radiation_series}), and (\ref{rescaled_a}) are solved iteratively in the same fashion. This eventually produces a configuration with the hydrostationary structure that is influenced by radiation and mass accretion.

A subtle point in this procedure is that the appropriate values of constants $\tilde C$ and $\tilde K$, as well as a multiplicative constant appearing in the rotation law, are not known a priori. One has to perform a kind of renormalisation of these constants during the iterative process described above.
From the condition that the inner and outer edges of the disk are located at $R_\mathrm{in}$ and $R_\mathrm{out}$, respectively (i.e., the density must vanish there), we can compute the values of $\tilde C$ and the multiplicative constant in the rotation law, provided that the potentials $\tilde \Phi$ and $\tilde \Psi$ are known. Expressing the rotation law as $\tilde \omega(\tilde r) = \bar \omega f(\tilde r)$ with $\bar \omega$ denoting the appropriate multiplicative constant, we can write
\begin{eqnarray}
\bar \omega^2 & = & \frac{1}{\int_{R_\mathrm{in}/R_\mathrm{out}}^{1} dr r f^2(r)} \left( \tilde \Phi(\tilde r = 1, \tilde z = 0) - \tilde \Psi(\tilde r = 1, \tilde z = 0) \right. \nonumber \\
& & \left. - \tilde \Phi(\tilde r = R_\mathrm{in}/R_\mathrm{out}, \tilde z = 0) + \tilde \Psi(\tilde r = R_\mathrm{in}/R_\mathrm{out}, \tilde z = 0) \right).
\label{rotation_const}
\end{eqnarray}
The value of $\tilde C$ can be now obtained from Eq.~(\ref{rescaled_a}) taken at $R_\mathrm{in}$ or $R_\mathrm{out}$ with $\tilde \rho = 0$.
The constant $\tilde K$ is given a value that ensures that the density corresponding to the found  maximal value of enthalpy $h$ is precisely $\rho_\mathrm{max}$. These computations have to be repeated each time a new density distribution is computed.

Convergence properties of our numerical scheme are dependent on the spatial resolution of the numerical grid. In this paper we follow the optimisation techniques for the computation of integrals appearing in Eqs.~(\ref{phi_series}) and (\ref{radiation_series}) that are described by \citet{mueller_steinmetz_95}. They allow one to achieve spatial resolutions of the order of $5000 \times 5000$ on a parallel computer consisting of 64 processors, with a total computing time of several minutes only. We truncate the expansion in Legendre polynomials $P_l$ in Eqs.~(\ref{phi_series}) and (\ref{radiation_series}) around $l_\mathrm{max} = 20$. The convergence is mainly controlled by computing the maximum norm of the two subsequent density distributions in the iteration scheme, i.e., we compute $\epsilon = \left\Vert \tilde \rho^{i+1} - \tilde \rho^{i} \right\Vert_\mathrm{max}$, where the index $i$ numbers subsequent iterations. We continue to iterate until $\epsilon$ reaches the level of $10^{-6},$ or even $10^{-8}$. Convergence in $\tilde \Phi$ and $\tilde \Psi$ is controlled in a similar fashion, the difference being that the maximum absolute values of these quantities are not known a priori.

\subsection{Numerical examples}

\begin{figure}
\begin{center}
\includegraphics[width=9cm]{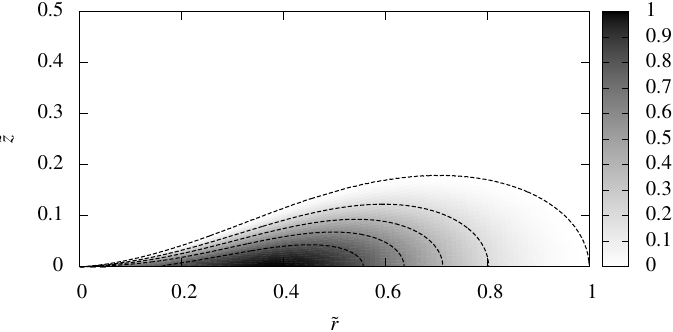}
\end{center}
\caption{Density $\tilde \rho$ obtained for solution $(a)$. The plot shows a section of the upper hemisphere in a meridian plane. The density is greyscale-coded.}
\label{rozw_a}
\end{figure}

\begin{figure}
\begin{center}
\includegraphics[width=9cm]{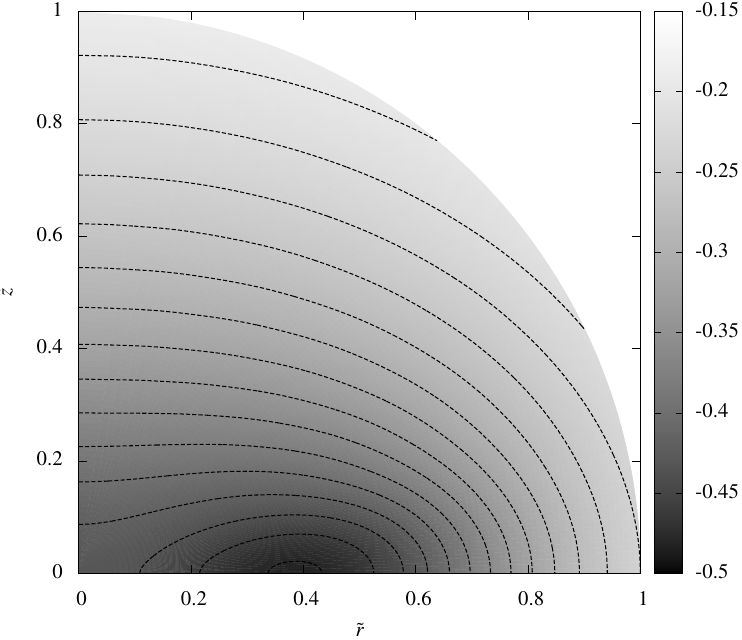}
\end{center}
\caption{Part of the gravitational potential $\tilde \Phi$ due to the accretion disk only, i.e., $\tilde \Phi + \tilde M_\mathrm{c}/\tilde R$, obtained for solution $(a)$. The plot shows a section of the upper hemisphere in a meridian plane.}
\label{phi_a}
\end{figure}

\begin{figure}
\begin{center}
\includegraphics[width=9cm]{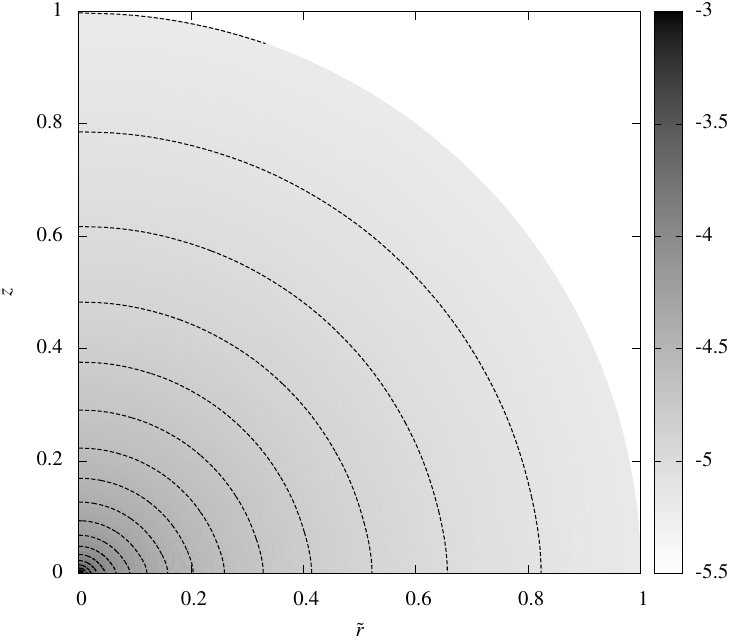}
\end{center}
\caption{Radiation potential obtained for solution $(a)$. The quantity $\log_{10} |\tilde \Psi|$ is plotted in greyscale. The graph shows a section of the upper hemisphere in a meridian plane.}
\label{psi_a}
\end{figure}

We discuss two sample solutions $(a)$ and $(b)$, obtained with the method described in the preceding section. They are computed for the polytropic exponent $\Gamma = 5/3$, and the central mass $\tilde M_\mathrm{c} = 2$. We assume the quasi-Keplerian rotation law $\tilde \omega = \bar \omega / \tilde r^{3/2}$. We refer to this rotation curve as ``quasi-Keplerian'' because $\bar \omega \neq \sqrt{\tilde M_\mathrm{c}}$, although the actual difference between $\bar \omega$ and $\sqrt{\tilde M_\mathrm{c}}$ appears small. Notice, for the subsequent discussion, that  setting $\bar \omega = \sqrt{\tilde M_\mathrm{c}}$ is equivalent to $\omega(r) = \sqrt{GM_\mathrm{c}}/r^{3/2}$. We take $\kappa \dot M$ distribution in the form
\begin{equation}
\kappa \dot M = A \exp\left( - \left( 50 \; \tilde z \right)^2 \right).
\label{accretion_f}
\end{equation}
Solution $(a)$ is obtained for $R_\mathrm{in}/R_\mathrm{out} = 10^{-4}$ and $A = 10^{-4}$. Solution $(b)$ is computed assuming $R_\mathrm{in}/R_\mathrm{out} = 10^{-3}$ and $A = 10^{-2}$. The density $\tilde \rho$, gravitational potential of the disk $\tilde \Phi + \tilde M_\mathrm{c}/\tilde R$, and the radiation potential $\tilde \Psi$, obtained for solution $(a)$ are shown in Figs.~\ref{rozw_a}, \ref{phi_a} and \ref{psi_a}, respectively. We do not display solution $(b)$, because its shape is not much different from that of solution $(a)$.

The obtained value of $\bar \omega$ is almost that of a strictly Keplerian rotation law. Expressing $\bar \omega$ as $\bar \omega = (1 + \alpha)  \sqrt{\tilde M_\mathrm{c}}$ we obtain $\alpha \approx 4.7 \cdot 10^{-6}$ for solution $(a)$ and $\alpha \approx 4.0 \cdot 10^{-5}$ for solution $(b)$. This can be intuitively understood by the careful inspection of Eq.~(\ref{rotation_const}); for elongated disks (i.e., for  low values of $R_\mathrm{in}/R_\mathrm{out}$),  the total gravitational potential is dominated by the divergent term $-G M_\mathrm{c}/R$  near the centre.

Other physical parameters of the solution depend on the choice of $R_\mathrm{out}$ and $\rho_\mathrm{max}$, as well as on the actual values of constants such as $\kappa$ and $G$. Possible values of $R_\mathrm{out}$ and $\rho_\mathrm{max}$ are restricted by the assumptions that were made when deriving equations of the model. The assumption of the thin-gas approximation requires that the mean free path of the photon $\lambda = m_\mathrm{p}/(\sigma \rho) = (c \kappa \rho)^{-1}$ should be comparable with the size of the entire configuration. The radial velocity $|U|$ should be lower than the angular one $r \omega$. Finally, the temperature of the disk should be high enough so that the assumption that the gas is fully ionised is justified. The radial velocity $U$ can be computed from Eq.~(\ref{bt}) based on the density distribution and the assumed mass accretion rate $\dot M$. Assuming the perfect gas approximation and a value of the mean molecular weight of the gas $\mu$, we can also find the distribution of the gas temperature
\[ T = p \mu m_\mathrm{p} / (\rho k_\mathrm{B}) = K \mu m_\mathrm{p} \rho^{\Gamma - 1} / k_\mathrm{B},\]
where $k_\mathrm{B}$ denotes the Boltzmann constant. For the fully ionised hydrogen one can take $\mu = 1/2$. The maximal temperature in the disk can be computed as
\[ T_\mathrm{max} = u \mu m_\mathrm{p} \tilde K /k_\mathrm{B}. \]

We take $R_\mathrm{out} = 10 \; \mathrm{parsecs}$ for both solutions, and assume $\rho_\mathrm{max} = 10^{-18} \mathrm{g \cdot cm^{-3}}$ and $\rho_\mathrm{max} = 10^{-17} \mathrm{g \cdot cm^{-3}}$ for solutions $(a)$ and $(b)$, respectively. This gives the central mass $M_\mathrm{c} = 2.9544 \cdot 10^7 M_{\astrosun}$ for solution $(a)$. The central mass for solution $(b)$ is higher by a factor of 10. Thus, if solutions $(a)$ and $(b)$ were to be applied in an astrophysical context, they would serve as models of accretion disks around ultramassive galactic black holes.

Other physical quantities characterising solutions $(a)$ and $(b)$ are as follows: For solution $(a)$: $M_\textrm{fluid} = (3.18720 \pm 0.00049) \cdot 10^6 M_{\astrosun}$, $L = (3.1101 \pm 0.0092) \cdot 10^6 L_{\astrosun}$, $T_\mathrm{max} = (3.73508 \pm 0.00031) \cdot 10^4 K$. For solution $(b)$: $M_\textrm{fluid} = (3.20779 \pm 0.00056) \cdot 10^7 M_{\astrosun}$, $L = (2.84443 \pm 0.00088) \cdot 10^9 L_{\astrosun}$, $T_\mathrm{max} = (3.72160 \pm 0.00036) \cdot 10^5 \mathrm{K}$. The error estimates given here were obtained by comparing solutions computed on numerical grids of different resolution (spanning the range of $2000 \times 2000$ to $4000 \times 4000$), with different numbers of the Legendre polynomials ($l_\mathrm{max} = 16$ to $l_\mathrm{max} = 22$), and different convergence level ($\epsilon = 10^{-6}$ up to $\epsilon = 10^{-8}$).

The estimate of the mean free path of a photon $\lambda$ -- evaluated for the maximal gas density -- is of the order of 1 parsec for solution $(a)$, which is roughly the maximal vertical height of the disk. For solution $(b)$ this estimate is 10 times smaller, but it is clear from the shape of the disk displayed on Fig.~\ref{rozw_a} that the optical thicknes is smaller than one, i.e.,
\[ \tau = \sup_{r \in [R_\mathrm{in},R_\mathrm{out}]} \int_{0}^{z_0} \frac{\sigma \rho(r,z)}{m_\mathrm{p}} dz \lesssim 1. \]
Here $z_0 = z_0(r)$ is the disk height at the cylinder of radius $r$ \citep{Mihalas}.
 
Numerical correctness of our solutions can be tested using the following virial theorem. Define $E_\mathrm{pot} = \int d^3 x \rho (\Phi + \Phi_\mathrm{Kep})/2$, $E_\mathrm{kin} = \int d^3 x \rho | \mathbf U |^2/2$, $E_\mathrm{therm} = \int d^3 x 3 p /2$, and $\tilde E = \kappa \int d^3 x \rho \mathbf x \cdot \mathbf j$, where $\Phi_\mathrm{Kep} = -G M_\mathrm{c}/R$ is the Keplerian potential of the point mass only (note that $\Phi_\mathrm{Kep}$ enters the formula for $E_\mathrm{pot}$ twice -- $\Phi$ is the total gravitational potential, i.e., the sum of $\Phi_\mathrm{Kep}$ and the contribution from the accretion disk). It can be shown \citep{mach_virial} that for a stationary configuration consisting of a point mass $M_\mathrm{c}$ and the fluid satisfying the Euler equation (\ref{aa}) the following virial identity holds:
\[ E_\mathrm{pot} + 2 E_\mathrm{kin} + 2 E_\mathrm{therm} + \tilde E = 0. \]
It differs from a standard formulation of the virial theorem by the presence of the point mass term in $E_\mathrm{pot}$, and the radiation coupling term $\tilde E$. The virial check of a numerical solution can be performed by computing 
\begin{equation}
\label{virial_th}
\epsilon_\mathrm{v} = |E_\mathrm{pot} + 2 E_\mathrm{kin} + 2 E_\mathrm{therm} + \tilde E|/|E_\mathrm{pot}|.
\end{equation}

We have obtained $\epsilon_\mathrm{v} \approx 10^{-8}$ for all numerical solutions. All constituent terms appearing in (\ref{virial_th}) were always much greater than this value. The following quantities are roughly the same for both solutions $(a)$ and $(b)$: $E_\mathrm{kin}/|E_\mathrm{pot}| \approx 0.478$, $E_\mathrm{term}/|E_\mathrm{pot}| \approx 2.22 \cdot 10^{-2}$, $\left(\int d^3 x \rho \Phi/2 \right)/E_\mathrm{pot} \approx 0.525$, $\left(\int d^3 x \rho \Phi_\mathrm{Kep}/2\right)/E_\mathrm{pot} \approx 0.475$. The ratio $\tilde E/|E_\mathrm{pot}|$ equals approximately $3.08 \cdot 10^{-6}$ for solution $(a)$ and $2.81\cdot 10^{-4}$ for solution $(b)$. That confirms the validity of our results. Clearly, our solutions pass the virial test surprisingly well. That is not very common, but not exceptional either \citep[see e.g.][]{aksenov_blinnikow}.

\section{Summary}

We have formulated a consistent model of a selfgravitating disk with steadily accreting matter. The radiation emitted by the disk interacts with the  gas by Thompson scattering (thin-gas approximation). The most interesting feature of our solutions is that the accretion mass rate flux density is concentrated in the equatorial plane $z=0$. The  slow radial drift of gas is observed in the equatorial plane and the bulk of gas remains approximately stationary. It appears that the conservation laws of the energy and the momentum together with the assumption of approximate stationarity suffice to obtain the structure of the disk. The emissivity index of accreting matter can be deduced after solving equations of the model. The approximation of stationarity demands that the radial inflow speed of matter has to be negligible compared with the rotational velocity of the gas in the disk. The secular change in the mass of the central accreting object should also be negligible. These two assumptions have been verified post factum for the numerical solutions presented in this paper.

The mathematical description of the radiating disk reduces to a pair of elliptic partial differential equations. They can be solved iteratively in a way similar to that routinely used in the literature when finding selfgravitating equilibria of non-radiating gas \citep[cf.][]{ostriker_mark_1968, eriguchi_muller_1985, nishida_eriguchi_lanza_1992}. In our case each iteration step consisted of finding new distributions of the density, the gravitational potential and the radiation potential. One of the main results of this paper is that this procedure numerically converges. Analytic results can be obtained in simplified cases. We investigated the influence of the emitted radiation onto the disk structure in the test fluid approximation. The interesting conclusion is that approximately stationary solutions do exist only when the mass accretion (and thus the luminosity) is not too large. This intuitively well-understood feature of solutions has been revealed also in our numerical analysis of heavy selfgravitating disks.

We assumed thin-gas approximation, slow radial flow of matter, and approximate stationarity. These conditions appear quite restrictive -- in the parameter space of solutions there is only an island of parameters for which the above assumptions can be satisfied. They lead to solutions with essential features that agree with the conditions encountered in some AGNs. We found solutions with the central mass of the order of $10^7 M_{\astrosun}$ to $10^8 M_{\astrosun}$ and disk masses of the order of $10^6 M_{\astrosun}$ to $10^7 M_{\astrosun}$. The luminosity varies between $10^6 L_{\astrosun}$ and $10^9 L_{\astrosun}$. We assumed the quasi-Keplerian rotation law $\omega \sim r^{-3/2}$, but the rotation appears in fact to be Keplerian in the examples described in this paper.

It is interesting that the luminosity in numerical examples has been much lower than the Eddington limit. This differs from the results of \citet{paczynski_abramowicz_1982} or \citet{hashimoto_et_al_1995}, where supercritical luminosity has been discovered. This discrepancy has a physical explanation -- we describe disks in the thin gas approximation, while the quoted works dealt with the disks in thermal equilibrium or with convection transport.

A future investigation of our model can be aimed in two directions: the study of stability of disks and the formulation of a general-relativistic version. Unlike for the standard models, the structure of our Newtonian disks is dependent on the accretion flux and radiation. We have discovered that Bondi-type, spherically symmetric solutions are stable also in the selfgravitating regime \citep{MM08}. The Bondi accretion models are spherically symmetric in contrast to accreting disks, but they share with our model the property that their structure depends on the accretion, and that can have a stabilising effect also in the nonspherical case. Thus we expect that the solutions discussed in this paper are stable.

There are two extreme classes of general-relativistic radiating  accretion disks.  Disks characterised by the size of an inner boundary exceeding $2 G M_\mathrm{c}/c^2$ (quasi-Newtonian case), where $M_\mathrm{c}$ is the central mass, should have a structure similar to those presented in this paper. However, their stability properties can still be different owing to the influence of gravitational radiation. Disks that are inherently relativistic, with the inner boundary located close to the minimum stable surface ($6 G M_\mathrm{c} /c^2$ for the Schwarzschild solution), can be significantly brighter than their Newtonian counterparts -- again, this expectation is based on our experience with Bondi-type solutions \citep{PRD2006, Rembiasz}.  

\begin{acknowledgements}
The research was carried out with the supercomputer ``Deszno'' purchased thanks to the financial support of the European Regional Development Fund in the framework of the Polish Innovation Economy Operational Program (contract no. POIG. 02.01.00-12-023/08).
\end{acknowledgements}


\begin{thebibliography}{}
\bibitem[Abramowicz, Calvani \& Nobili(1980)]{abramowicz_calvani_nobili_1980} Abramowicz, M.A., Calvani, M., \& Nobili, L. 1980, ApJ, 242, 772
\bibitem[Abramowicz, Jaroszy\'{n}ski \& Sikora(1978)]{abramowicz_jaroszynski_sikora_1978} Abramowicz, M.A., Jaroszy\'{n}ski, M., Sikora, M. 1978, A\&A, 63, 221
\bibitem[Axenov \& Blinnikov(1994)]{aksenov_blinnikow} Axenov, A.G., \& Blinnikov, S.I. 1994, A\&A 290, 674
\bibitem[Blinnikov(1975)]{blinnikov_1975} Blinnikov, S.I. 1975, Soviet Astron. 19, 151
\bibitem[Bondi(1952)]{Bondi} Bondi, H. 1952, MNRAS, 112, 195
\bibitem[Clement(1974)]{clement_1974} Clement, M.J. 1974, ApJ, 194, 709
\bibitem[Eriguchi(1978)]{eriguchi_1978} Eriguchi, Y. 1978, Publ. Astron. Soc. Japan, 30, 507
\bibitem[Eriguchi \& M\"{u}ller(1985)]{eriguchi_muller_1985} Eriguchi, Y., \& M\"{u}ller, E. 1985, A\&A, 146, 260
\bibitem[Hashimoto et.~al(1993)]{hashimoto_et_al_1993} Hashimoto, M., Eriguchi, Y., Arai, K., \& M\"{u}ller, E. 1993, A\&A, 268, 131
\bibitem[Hashimoto et.~al(1995)]{hashimoto_et_al_1995} Hashimoto, M., Eriguchi, Y., \& M\"{u}ller, E., 1995, A\&A, 297, 135
\bibitem[Karkowski et al.(2006)]{PRD2006} Karkowski, J., Kinasiewicz, B., Mach, P., Malec, E., \& Swierczy\'{n}ski, Z. 2006, Phys. Rev. D, 73, 021503(R)
\bibitem[Komatsu, Eriguchi \& Hachisu(1989)]{komatsu_eriguchi_hachisu_1989} Komatsu, H., Eriguchi, Y., \& Hachisu, I. 1989, MNRAS, 237, 355
\bibitem[Lynden-Bell \& Pringle(1974)]{LBP} Lynden-Bell, D., \& Pringle, J.E. 1974, MNRAS, 168, 603
\bibitem[Mach(2012)]{mach_virial} Mach, P. 2012, MNRAS, 422, 772
\bibitem[Mach \& Malec(2008)]{MM08} Mach, P., \& Malec, E. 2008, Phys. Rev. D, 78, 124016
\bibitem[Malec \& Rembiasz(2010)]{Rembiasz} Malec, E., \& Rembiasz, T. 2010, Phys. Rev. D, 82, 124005
\bibitem[Mihalas \& Weibel-Mihalas(1984)]{Mihalas} Mihalas, D. \& Weibel-Mihalas, B. 1984, Foundations of Radiation Hydrodynamics, (Oxford University Press)
\bibitem[M\"{u}ller \& Steinmetz(1995)]{mueller_steinmetz_95} M\"{u}ller, E. \& Steinmetz, M. 1995, Comput. Phys. Commun. 89, 45
\bibitem[Nishida, Eriguchi \& Lanza(1992)]{nishida_eriguchi_lanza_1992} Nishida, S., Eriguchi, Y., \& Lanza, A. 1992, ApJ, 401, 618
\bibitem[Odrzywo\l ek(2003)]{Odrzywolek} Odrzywo\l ek, A. 2003, MNRAS 345, 497
\bibitem[Ostriker \& Mark(1968)]{ostriker_mark_1968} Ostriker, J.P., \& Mark, J.W-K. 1968, ApJ, 151, 1075
\bibitem[Paczy\'{n}ski(1978)]{Paczynski78} Paczy\'nski, B. 1978, Acta Astronomica, 28, 91
\bibitem[Paczy\'{n}ski \& Abramowicz(1982)]{paczynski_abramowicz_1982} Paczy\'{n}ski, B., \& Abramowicz, M.A. 1982, ApJ, 253, 897 
\bibitem[Paczy\'{n}ski \& Wiita(1980)]{Wiita} Paczy\'nski, B., \& and Wiita, P.J. 1980, A\&A, 88, 23
\bibitem[Padmanabhan(2000)]{padmanabhan} Padmanabhan, T. 2000, Theoretical Astrophysics, Vol.~I, (Cambridge Univ. Press)
\bibitem[Papaloizu \& Pringle(1985)]{papaloizu_pringle_85} Papaloizu, J.C.B., \& Pringle, J.E. 1985, MNRAS, 213, 799
\bibitem[Papaloizu \& Pringle(1984)]{papaloizu_pringle_84} Papaloizu, J.C.B., \& Pringle, J.E. 1984, MNRAS, 208, 721
\bibitem[Piran(1978)]{piran_78} Piran, T. 1978, ApJ, 221, 652
\bibitem[Pringle(1976)]{pringle_76} Pringle, J.E. 1976, MNRAS, 177, 65
\bibitem[Pringle(1981)]{pringle_81} Pringle, J.E. 1976, ARAA, 19, 137
\bibitem[Qian et al.(2009)]{Qian} Qian, L., Abramowicz, M.A., Fragile, P.C., Hork, J., Machida, M., \& Straub, O. 2009, A\&A, 498, 471Q	
\bibitem[Shakura \& Sunyaev(1973)]{ShSu} Shakura, N.I., \& Sunyaev, R.A. 1973, A\&A, 24, 337
\bibitem[Shakura \& Sunyaev(1976)]{shakura_sunyaev_76} Shakura, N.I., \& Sunyaev, R.A. 1976, MNRAS, 175, 613
\bibitem[Sikora(1981)]{Sikora} Sikora, M. 1981, MNRAS, 196, 257
\bibitem[Stoeckly(1965)]{stoeckly_1965} Stoeckly, R. 1965, ApJ, 142, 208

\end{thebibliography}
\end{document}